\documentclass[letterpaper,11pt,preprint]{aastex}
\usepackage{graphics,graphicx}
\usepackage{natbib}

\usepackage{color}
\makeatletter
\renewcommand{\section}{\@startsection%                                                                                                                                                                    
{section}{1}{0mm}{-\baselineskip}%                                                                                                                                                                         
{0.5\baselineskip}{\normalfont\Large\bfseries}}%                                                                                                                                                           
\makeatother

\newcommand{\one}{{\mathbf 1}}

\begin{document}
\pagestyle{plain}

\title{Calibration of Quasi-Redundant Interferometers}

\author{J. L. Sievers\altaffilmark{1,2}}

\altaffiltext{1}{School of Chemistry and Physics, University of KwaZulu-Natal, Private Bag X54001, Durban 4000, South Africa}
\altaffiltext{2}{National Institute for Theoretical Physics (NITheP), KZN node, Durban 4001, South Africa}

\begin{abstract}
High precison calibration is essential for a new generation of radio
interferometers looking for Epoch of Reionization and Baryon Acoustic
Oscillation signatures in neutral hydrogen.  These arrays have so far
been calibrated by redundant calibration, which usually assumes baselines
intended to be identical are perfectly so.  We present a new
calibration scheme that relaxes the assumption of explicit redundancy
by calculating the expected covariance of baselines.  The technique
also allows one to take advantage of partial knowledge of the sky,
such as point sources with known positions but unknown fluxes.  We
describe a 2-level sparse matrix inverse to make the calibration
tractable for 1,000-element class interferometers.  We provide a
reference implementation and use it to test the calibration of
simulations of an array with imperfectly located antennas observing
Euclidean-distributed point sources.  Including position information
for a handful of the brightest sources, we find the amplitude/phase
reconstruction improves by a factor of $\sim$2/5 over redundant
calibration for the noise levels/position errors adopted in the
simulations.  Inclusion of source positions also allows us to measure
the overall phase gradient across the array, information which is lost
in traditional redundant calibration.

\end{abstract}

{\noindent
\section{Introduction}

Several current and near-future radio arrays, such as PAPER \citep{paper8,paper64},
  HERA \citep{hera}, CHIME \citep{chime_pathfinder,newburgh2014}, Tianlai
  \citep{tianlai}, and HIRAX \citep{hirax}, will consist primarily of
close-packed detecting elements carrying out drift scans, with highly
redundant baseline distributions.  Calibrating these arrays presents a
challenge, as the sky signal is typically dominated by unkown diffuse
emission. This present challenges for traditional methods such as
selfcal \citep{selfcal}, since the notion of a common sky model seen
by all baselines, while formally true, ceases to have relevance.  The
calibration problems for these experiments is particularly acute since
they are all searching either for Epoch of Reionization (EoR) signals
or Baryon Acoustic Oscillations (BAOs); for both science cases the
foreground to signal ration is often $10^3$ or greater.  Developing a
sky model of sufficient quality for traditional calibration may be
impossible in practice for these use cases \citep{skacal}.  For
redundant arrays, an alternate approach (Omnical, \citet{omnical}) is
to rely on the fact that all redundant baselines should see the same
signal.  With this assumption, Omnical can then solve for the true sky
and the relative antenna gains by solving an over-determined linear
least-squares problem.  This procedure has been successful in
calibrating redundant arrays such as PAPER.  However, one shortcoming
is that in real life, arrays are never perfectly redundant due to
imperfect antenna locations and variations in primary beams.  In this
work, we present an alternative scheme to carry out the relative
calibration that can take account of array imperfections.  The method
also generalizes to naturally include bright sources with known
positions while still allowing the diffuse emission to be the dominant
signal.
 
\section {Likelihood Formulation}

\subsection{Traditional Redundant Calibration}
In traditional redundant calibration, the instrument is modelled as a
set of per-antenna (complex) gains, and a set of unknown sky values,
defined at a finite set of points in the UV plane.  All visibilities
at a given UV point are expected to see the same sky value.  Under
these assumptions, the predicted measured visibility between antennas $i$ and
$j$ is 
$$g_i^*g_j S(u_i-u_j)$$
for antenna gains $g_i$ and $g_j$ and sky brightness $S(u_i-u_j)$ at
the UV point corresponding to the vector spacing between antennas $i$
and $j$, measured i wavelengths.  

The form of $\chi^2$, under the assumption of noise uncorrelated
between visibilities, is then:
$$\chi^2=\sum \frac{(v_{ij}-g_i^*g_jS(u_i-u_j))^2}{\sigma_{ij}^2}$$ It
is relatively straightforward to take the gradient of the likelihood
and then find the global solution for the sky and gains that minimizes
$\chi^2$ ({\textit{e.g.}} \citet{ram_marthi_chengalur}).  If there are
enough redundant baselines so that the number of visibilities exceeds
the number of antennas plus the number of unique baselines, then the
solution is well determined, up to four degeneracies built into the
likelihood.  If one multiplies the gains by some factor $\alpha$, and
divides all sky values by $\alpha^2$, then the $\alpha$'s cancel, and
$\chi^2$ is explicitly unchanged.  So, redundant calibration cannot
set the overall gain of the instrument.  Similarly, a global phase
shift produces identically unchanged visibilities.  Third, one can
also apply phase gradients to the gains in the $x-$ and
$y-$directions, and apply opposite phase gradients to the sky, and
again leave $\chi^2$ explicitly unchanged.  This is equivalent to
changing the pointing center of the array, then shifting the sky model
to the new pointing center.

%This expression for $\chi^2$ can
%be re-written as:
%$$\chi^2=(d-g_1^*g_2S)\rm{N}^-1(d-g_1^*g_2S)$$.

Traditional redundant calibration suffers from two key assumptions
than can often be improved upon in realistic cases.  First, solving
for the sky implicitly marginalizes over all possible skies, which are
considered equally likely.  We generally don't have perfect sky
models, but we often have {\textit{some}} idea of what the sky looks
like - for instance, the position of bright sources is often known,
even if their fluxes at the frequency and time of observation aren't.
Second, traditional redundant calibration assumes the array is
explicitly redundant, and including information about a non-ideal
array is rather awkward in the formalism.  For instance, \citet{omnical} include
UV-plane gradients in the solution as first-order corrections for an
imperfect array.  This solution however suffers from several
shortcomings.  First, two nearly redundant baselines should see nearly
the same sky - by solving for gradients, one effectively says that
they can be arbitrarily different, which unnecessarily increases the
noise in the recovered calibration, and in extreme cases can cause the
system to become ill-conditioned.  Second, realistic arrays will
generally have properties that do not correspond to simple gradients
in the UV plane.  For instance, even if the dishes are all pointed in
the same direction, small variations in dish manufacturing will lead
to small variations in the primary beams.  For a realistic array with
scatters in pointing, dish positioning, and primary beams, expanding
traditional redundant calibration to include the full range of
possible effects rapidly becomes intractable.

\subsection{Correlation Calibration}
We now describe a new calibration scheme that has the advantages of
traditional redundant calibration (and in fact contains traditional
redundant calibration as one limiting case) while also capable of
flexibly including more realitic models for the instrument and sky.

While there are several ways to derive the formalism, one way is to
rewrite the redundant formalism, gradually relaxing its assumptions.
The key step is to remove the explicit dependence on the sky that
redundant calibration has.  In redundant calibration, if one specifies
the antenna gains, then it is straightforward to solve for the sky
values. They are just the (inverse variance-weighted) averages of the
visibilities at each unique UV point.  Now, consider the block of
visibilities from a single unique UV point.  Instead of solving for
the sky value, one can add the unknown sky into the noise for this
block.  They all see the same sky value $\alpha$, so the covariance
between two visibilities from the sky is $\left < V_i^*V_j\right >$=$\alpha^*\alpha$.
Recall that in the absence of signal and in the presence of noise
correlated between data points, the form of $\chi^2$ is
$$\chi^2=d^\dag \mathrm{N}^{-1}d$$
where $<d_i^{\dag}d_j>=N_{ij}$.
For a redundant block, the effective noise can be written as the sum
of two terms:  a diagonal matrix of per-visibility noises, and a
matrix that is the outer product of the vector $\alpha$ times a vector
of ones with itself.  
$$N=N_{vis}+(\alpha \one)^\dag (\alpha \one)$$.  
This matrix can be inverted using the Woodbury identity, which we
write here in the special case of a Hermitian matrix:
\begin{equation}
(A+bb^{\dag})^{-1}=A^{-1}-A^{-1}b(I+b^\dag A^{-1} b)^{-1}
  b^{\dag}A^{-1}
\label{eq:woodbury}
\end{equation}
.  Now let $A=N_{vis}$ and $b =\alpha\one$, and define weight matrix
$W\equiv N_{vis}^{-1}$.  The full form of $\chi^2$ is now:
$$\chi^2=d^\dag \left ( W-W\one (\alpha^{-2}+\one^\dag W \one )^{-1}
\one^{\dag} W \right ) d$$ The term $\beta \equiv \one^\dag W d$ is
just the weighted sum of the data.  The term $\gamma \equiv \one^\dag
W \one$ is just the sum of the weights.  So, our best estimate of the
sky is just the weighted average of the data, or
$\frac{\beta}{\gamma}$.  If we take the limit of infinite sky
variance, or $\alpha \rightarrow \infty$ (which forces the prior
probability of all possible skies to be equal), then we have
$$\chi^2=d^\dag W d - \frac{\beta^2}{\gamma}$$
where $\beta^2$ is understood to be the squared magnitude of the
complex number $\beta$, and since both $\beta$ and $\gamma$ are
scalars, multiplication commutes and inverses are just divides.  

Let us now compare this to the value of $\chi^2$ we would have gotten
by fitting for the sky.  Since the best fit sky is
$\frac{\beta}{\gamma}$, we can subtract that from the data vector and
calculate $\chi^2$ as usual:
$$\chi^2=(d-\frac{\beta}{\gamma}\one)^\dag W (d-\frac{\beta}{\gamma}\one)$$
$$=d^\dag W d -2\frac{\beta^\dag}{\gamma}\one^\dag W d +
\frac{\beta^\dag}{\gamma}\one^\dag W \one \frac{\beta}{\gamma}$$
where we have used the fact that all matrices are Hermitian to combine
the two cross terms.  Again, $\one^\dag W d=\beta$, and $\one^\dag W
\one=\gamma$, so we are left with 
$$\chi^2=d^\dag W
d-2\frac{\beta^2}{\gamma}+\frac{\beta^2}{\gamma}=d^\dag W
d-\frac{\beta^2}{\gamma}$$.  
This is identical to the expression for $\chi^2$ we obtained when
putting the sky in as a noise term, and so the two methods
{\textit{must}} be equivalent.  

\subsection{Relaxing Redundancy}

Switching to a covariance-based formulation of $\chi^2$ provides
significant flexibility.  If we back away from the infinite signal
$\alpha \rightarrow \infty$ limit, and instead treat the sky as a
Gaussian random field described by a spatial power spectrum, then
covariances between non-identical baselines can be calculated.  There
is a long literature, particularly from Cosmic Microwave Background
experiments, on how to do this
\citep{white1999,hobson_maisinger,cbigridr}.  Essentially, one
integrates the product of the UV-plane primary beams of two baselines,
weighting by the power spectrum as a function of $|u|$.  Pointing
offsets can be described by phase gradients in the UV-space primary
beam, mis-positioned dishes shift the UV centers of the primary beams,
and dish non-uniformities change the shape of the primary beam.  So,
the covariances from a wide variety of non-idealities can be
calculated relatively straightforwardly.  Of course, the sky is not
actually a Gaussian random field, but the assumption should be quite
robust for realistic skies - wee the Appendix for a discussion.

Furthermore, sources with known positions can also be included, even
if the fluxes are not precisely known.  If the predicted visibilities
from a source are $q$, then we add $q^\dag q$ to the covariance
matrix.  This effectively tells the likelihood that there is less of a
penalty for putting signal where a source is known to exist than
spreading it out across the map.  If the expected amplitude is
unknown, one can simply take the limit as $q$ goes to infinity, which
fully marginalizes over the source flux.

\section{Minimizing $\chi^2$}

Naively, evaluating $\chi^2$ requires inverting a matrix with
dimension $n_{vis}$.  For 1,000 element-class instruments that have
$\sim 10^6$ visibilities, brute-force inversion is computationally
unfeasable.  With care, however, the sparseness of typical
quasi-redundant systems can be used to quickly solve for telescope
gains.  

Starting with $\chi^2$, we have:
$$\chi^2=(Gd)^\dag (G^\dag N G + C)^{-1} Gd$$
again, for data $d$, diagonal gain matrix $G$ where the $i^{th}$
diagonal entry is the product of the conjugate of the gain of the
first antenna with the gain of the second antenna, $N$ is the observed
visibility noise variance, and $C$ is the expected data covariance.
We assume the visiblity noises are uncorrelated with each other, which
is usually well-justified in radio astronomy.  This expression can be
simplified if we instead use the inverse gains $H\equiv G^{-1}$.  Then
we multiply through by $H^\dag$ on the left and $H$ on the right,
leaving:
\begin{equation}
\chi^2=d^\dag (N + H^\dag C H)^{-1}d
\label{eq:chisq}
\end{equation}
. 

\subsection{Evaluating $\chi^2$}

For computational efficiency, we restrict the form of $C$.  If we
group the visibilities into quasi-redundant blocks, then $C$ consists
of a set of blocks along the diagonal desribing the covariances within
a quasi-redundant block plus a term describing the contribution of
known sources.  We store the blocks and the source contributions as a
set of vectors whos outer products form the quasi-redundant/source
contributions to the signal covariance. 

The real-imaginary symmetry usually found in radio astronomy is broken
on short baselines \citep{cbigridr} and strongly broken by the
inclusion of point sources with fixed positions.  So, we switch to an explicitly real
formulation of $\chi^2$.  The inverse gain matrix $H$ now consists of
2x2 blocks along the diagonal, assuming the data vector is written
$[v_{1,r}\ v_{1,i}\ v_{2,r}\ v_{2,i} ...]$.  Call the matrix of source
vectors $S$ and the matrix of quasi-redundant vectors $R$.  For
perfectly redundant data, $R$ will consist of two vectors for each
redundant block, one corresponding to the real part of the
visibilities, and one to the imaginary part: $R_{r,:}=\alpha
[1\ 0\ 1\ 0...]$ and $R_{i,:}=\alpha [0\ 1\ 0\ 1...]$.  For
imperfectly redundant cases, $R$ for each block can be approximated by
sufficiently large eigenvalues and their corresponding eigenvectors
$\lambda_j^{1/2} v_j$.  If these blocks can {\textit{not}} be
accurately represented by many fewer eigenvectors than there are
visibilites within a block, then there is not sufficient redundancy in
the data and any calibration that relies on redundancy is unlikely to
provide a satisfactory solution.

With the sky/sources expressed as vector outer products, we now have 
$$\chi^2=d^T\left ( N+H^T(SS^T+RR^T)H \right )^{-1}d$$.  This matrix
can fortunately be efficiently inverted by a repeated application of
the Woodbury identity.  First, apply the inverse gains to the
source/redundant vectors, $\hat{S}=H^{T}S, \hat{R}=H^{T}R$.  Then,
invert each redundant block using the 
Woodbury identity while including visibility noise but ignoring
sources $\Gamma^{-1}$ where $\Gamma \equiv N+\hat{R}\hat{R}^{T}$.  Each block is kept separate in
the inverse, and saved in a factored form, typically by applying the
Cholesky factorization of the matrix $(I+b^\dag A^{-1} b)^{-1}$ from
Equation \ref{eq:woodbury} to the redundant vectors.  Finally, do a
final inverse using the source vectors:
\begin{equation}
  \left ( \Gamma + \hat{S}\hat{S}^{T} \right )^{-1}
\label{eq:inverse}
\end{equation}
where we already have the factored form of $\Gamma^{-1}$.  This makes the matrix-vector
operations for the second application of the Woodbury identity much
faster.  Implemented this way, the inverse is efficient, with
the operation count scaling like $n_{vis}$ times
$max(n_{src},n_{redundant})^2$.  For 1,000 antennas, 1 source and 2
redundant vectors, a single core on a laptop-class machine can
evaluate $\chi^2$ this way in $\sim 0.1$ seconds using a mixed
C-Python code.

\subsection{Minimizing $\chi^2$}

While there are several possible algorithms to find the best-fit
gains, we implement a non-linear conjugate-gradient solver using
gradient information.  Just inverting a curvature matrix would scale
like $n_{ant}^3$ (let along the work involved in calculating it), so
the $\chi^2$ evaluation should be faster than that as long as
$max(n_{src},n_{redundant})$ is less than about $n_{ant}$.  The
gradient of $\chi^2$ with respect to antenna gains is:
$$\nabla \chi^2 = d^T (N+H^TCH)^{-1} (H^{'T}CH + H^TCH^{'})(N+H^TCH)^{-1} d$$ 
where $H^{'}$ is the derivative of $H$ with respect
to an antenna gain.  As usual, the two terms in the central block end
up numerically identical, so we have
$$\nabla \chi^2 = 2d^T (N+H^TCH)^{-1}  (H^{'T}CH) (N+H^TCH)^{-1} d$$.
To evaluate this, first form $p\equiv (N+H^TCH)^{-1} d$ then $q\equiv
CHp$.  These matrix-vector multiplies are very fast, and are the same
for all antenna gains.  This leaves us with 
\begin{equation}
\nabla \chi^2 = 2p^T H^{'T} q = 2q^T H^{'} p
\label{eq:chi_grad}
\end{equation}.  Each 2x2 block of $H$
only has the gains from 2 antennas in it, so rather than looping over
antennas, it is more efficient to loop over the visibilities, and
accumulate the per-antenna contribution to the gradient.  

\subsection{Reference Implementation}
We provide a reference Python/C implemenation available from github,
located at\\ https://github.com/sievers/corrcal2.  
For small ($\sim 100$ antenna)
problems with only a few source/redundant vectors, using BLAS/LAPACK actually
slows down the code, since the function overhead for {\textit{e.g.}}
Cholesky decomposition of 2x2 blocks is non-negligible.  So, the
reference implementation fully self-contained.  More
complicated cases will undoubtedly wish to link to higher-performance
libraries.  Minimization of $\chi^2$ is done using the Scipy \citep{scipy}
nonlinear conjugate-gradient solver.  For visibilties with similar
noises and starting from a reasonably close calibration (off by tens
of percent) we find for 1,000 antenna-class problems, including a
single source and a perfectly redundant, array that a single
laptop-class core can evaluate $\chi^2$ in less than $0.1$s, can evaluate
the gradient in $\sim 0.25$s, and can solve for a full calibration
solution in 15-25s.  

The current reference implementation requires only a C compiler plus
Python.  Future work will likely include optimizing the minimization
routine for typical calibration problems, emphasizing data re-use
since small problems are highly memory-limited, and a GPU port.  

\section{Application to Sample Data}\label{sims}

While we defer a full investigation of the quantitative behavior of
correlation calibration relative to redundant calibration to future
work, we present here results from a test run.  For the test run, we
take an 8x8 array on a quasi-regular grid with antennas spaced by 20
wavelengths with a random noise of 0.04 wavelengths in the antenna
positions in both the $x-$ and $y-$directions.  The primary beams are taken to be
Gausians and correspond to 13 wavelength-diameter dishes.  For
simplicity, we assume a flat-sky treatment; the results will not
change qualitatively for a curved-sky simulation. A total of 12,500
sources are randomly distributed within $2.5\sigma_{PB}$ with
Euclidean-distributed fluxes.  Sources with primary beam-weighted flux
greater than 3 times the RMS are treated as having known positions,
which typically results in $\sim 10$ individually-treated sources.
The remainder are treated by calculating the visibility covariance
within quasi-redundant blocks under the assumption of a Poisson power
spectrum.  Eigenmodes with amplitude more than $10^{-6}$ times the
largest are kept.  This results in keeping 2-3 complex eigenmodes per
quasi-redundant block. Since we are primarily interested in systematic
errors in calibration, we add only a small amount of per-visibility
noise - 0.1 times the bright source threshold.  The calibration is
then solved for three cases: 1) correlation calibration including
position information on the brightest sources, 2) correlation
calibration but not using bright source information, and 3) redundant
calibration.  The redundant calibration is actually carried out using
the correlation calibration pipeline, but with the covariances
calculated using the nominal grid positions rather than the actual
ones and the sky covariance multiplied by a large factor to
approximate the effects of sky marginalization.

We carry out roughly 2,000 sets of comparison simulations. The true
gains are exactly one in the simulation, but the starting calibration
fed to the conjugate-gradient solver is randomly offset by 20\% in both
the real and imaginary components.  Comparing the results between
methods is more subtle than might be expected, since the overall
normalization is poorly/unconstrained, and the phase gradient is
unconstrained except when bright sources are included.  So, comparison
to the true calibration is often not reflective of actual
performance.  We therefore
use the following statistics to compare methods.  For the amplitude,
we report the standard deviation of the calibration relative to the
mean, and for the phase we first fit an offset and gradient across
the array and then report the standard deviation relative to that.
Since the true calibration is unity, amplitude/phase correspond to the
real/imaginary parts of the solution to second order.

The simulation results, plotted in Figure \ref{sim_outputs}, are consistent with
expectations.  The median amplitude/phase standard deviations of the
full source/covariance treatment are $8.3\times 10^{-4}$ and $7.0
\times 10^{-4}$.  For the full covariance treatment when ignoring
bright source positions the standard deviations are $9.2 \times
10^{-4}$ and $4.2 \times 10^{-3}$.  Finally, the redundant case gives
$1.7 \times 10^{-3}$ and $4.4 \times 10^{-3}$.  These results show a
clear heirarchy.  When the true antenna positions are used, the
calibration amplitude is better - a factor of 2 for the antenna
noise/primary beams used here.  The bright source information only
provides a modest improvement on the amplitude calibration.  However,
the story is flipped for the phase calibration.  The source-aware
calibration has significantly lower phase error - for the simulation
parameters it is a factor of 5 better than either redundant
calibration or non-source aware correlation calibration.  The two
cases that do not include bright source information perform very
similarly on the phase calibration, with the correlation calibration
performing very slightly ($\sim$5\%) better than redundant
calibration.  

\begin{figure*}[thbp]
  \resizebox{0.55\hsize}{!}{\includegraphics{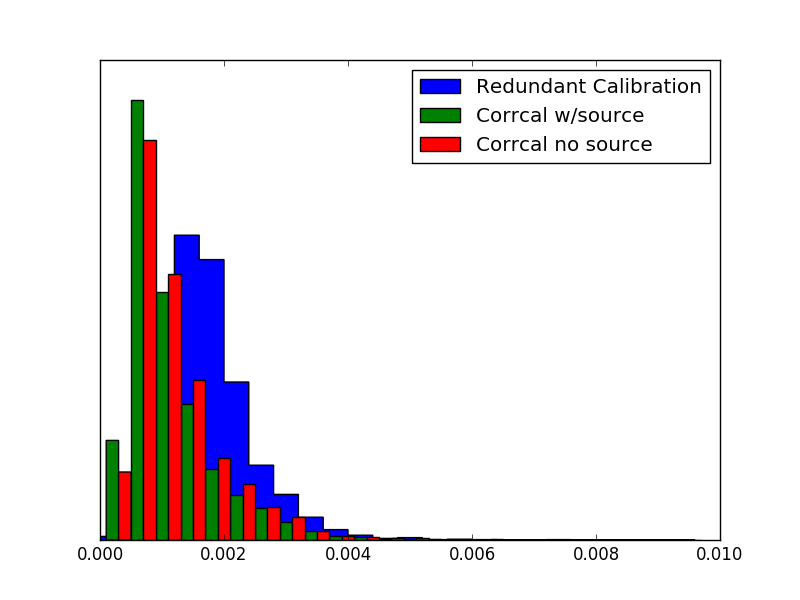}}
  \resizebox{0.55\hsize}{!}{\includegraphics{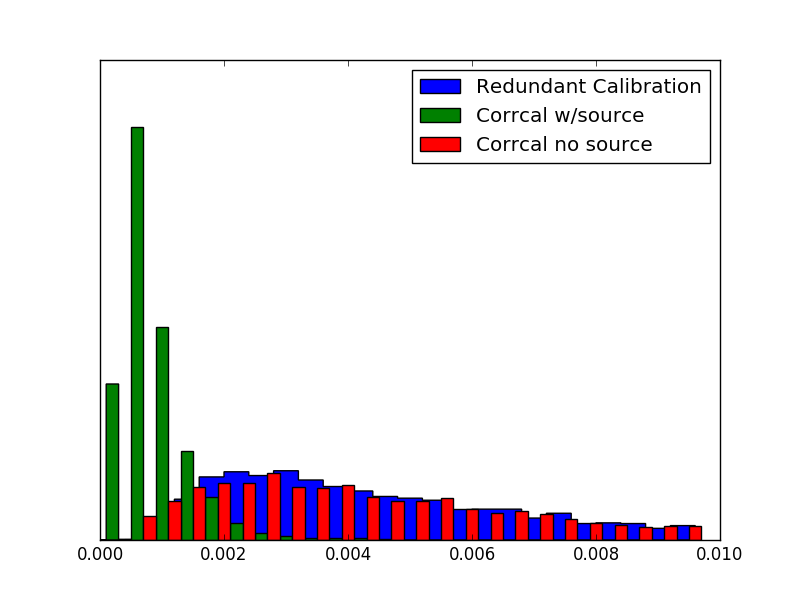}}\\
  \caption{Histograms of the scatter in recovered amplitude (left) and
    phase (right) for calibration on simulated source populations.
    The blue solid histograms shows the results from redundant
    calibration.  The green histograms show the results when the
    correct covariance given true antenna positions and knowledge of
    the positions of the brightest sources are used.  The red
    histograms use the correct covariances but do not use knowledge of
    source positions.  The red histograms have been offset from the
    green for clarity.  Using the correct covariances improves the
    amplitude reconstruction while using the source positions
    dramatically improves the phase reconstruction.  The antennas are
    spaced by 20 wavelength, have a Gaussian primary beam size
    equivalent to 13-wavelength dishes, and have positonal scatters of
  $0.04\times\sqrt(2)$.}
  \label{sim_outputs}
\end{figure*}

\section{Future Applications}
The formulation of $\chi^2$ presented here is highly general and can
apply to a wide range of situations.  We outline several possible use
cases that will be explored in further work. 

\subsection{Non-redundant Arrays}

At its heart, redundant calibration relies on the fact that there are
far more visibilities than instrumental plus sky degrees of freedom.
Radio telescopes that place antennas to maximize UV coverage will
likely not be calibratable this way, but arrays with sufficiently
dense cores should be.  There, a natural procedure is to break the UV
plane up into tiles, calculate covariance within a tile, and ignore
covariances between tiles.  This will be sub-optimal to the extent
that cross-tile covariances are non-negligible.  But, it should not be
biased - for instance classical redundant calibration ignores the
correlations from sources (which correlate across UV position) but is
unbiased.

\subsection{Solving for Antenna Properties}

The formulation presented here assumes known array properties.
However, a conceptually simple extension can allow it to be used to
estimate antenna properties.  The value for $\chi^2$ is properly
normalized, and so absolute goodness-of-fits can be compared for
varying antenna models.  

One way of implementing this would be to loop over antennas, running
Markov chains to solve for antenna positions/pointing offsets/beam
deformations.  For a small number of parameters, Markov chains
converge quickly, often in just a few hundred likelihood evaluations.
Re-diagonalizing blocks of the covariance matrix would be very
expensive, but when only a single antenna is having its properties
varied, the covariance matrix can be efficiently inverted using
a partitioned inverse at the expense of increased bookkeeping.  

The covariances between the beam-varied baselines and the reference
baselines must also be calculated.  Within a redundant block, the
non-varied antennas must have a sparse representation for redundancy
to hold - found {\textit{e.g.}} by looking at the SVDs of the
visibility beam maps.  Rather than calculate the baseline covariances
explicitly, a further speedup should be attainable by correlating
against the sparse representation, and then using that to calculate
the visibility covariances.  In the regime where the varied antenna
properties stay in the same sparse space, a further speedup can be
achieved by staying in that space where all operations are of order
$n_{mode}$.  

\subsection{Bandpass Calibration}
Bandpass calibration is particularly crucial for experiments searching
for faint signals in the presence of high foregrounds.  Ripples in the
calibration, either in amplitude or in phase, can allow foregrounds to
leak into EoR or BAO signals. Correlation calibration provides a
natural formalism to carry out cross-frequency calibration.  Sources
can {\textit{e.g.}} be described with a few modes, one corresponding
to the overall amplitude and one or more to uncertainties in the
spectral behavior.  Most simply this would be an uncertainty in the
spectral index, but it is straightforward to extend to include more
complicated spectral behavior such as curvature.  The cross-frequency
behavior of foregrounds can also be estimated under assumptions about
the spectral properties, and relevant eigenmodes included in the
calibration.  Combined, these extensions should naturally force the
resulting sky images to be spectrally smooth.  We note that the
overall spectral index will likely be poorly constrained since it is
effectively set by the assumed average index of the sky/sources, but
the higher order calibration should be robust.

\subsection{Polarization Calibration}

Correlation calibration can also be extended to cover polarization
calibration.  Including polarization angles will allow the calibration
to force a consistent view of the sky polarization between antennas.
In addition, one or more calibrators with known polarization can be
included by not giving a degree of freedom to the polarization angle.
This will lock the solution to the calibrator(s) while correctly
accounting for uncertainies from background diffuse emission.  

\section{Conclusions}
Arrays with nominally redundant antenna positions are becoming
increasingly common in astronomy.  They are particularly prevelant in
Baryon Acoustic Oscillation and Epoch of Reionization studies.  While
redundant calibration \citep{omnical} marked a great step forward,
real arrays will never be exactly redundant.  Calibration using a
Gaussian likelihood allows one to take array non-uniformities into
account, and in addition provides a natural framework to include
partial knowledge of the sky.  When implemented using a 2-level sparse
matrix description, the resulting likelihood is fast to calculate for
even thousand element-class arrays, and canned solver routines display good
convergence properties.  We apply correlation calibration to
simulations with antennas with small errors in their positioning and
find that calibration amplitude reconstruction is improved.  When
knowledge of the positions of the brightest sources is included, the
phase reconstruction improves significantly as well.

One of the strengths of correlation calibration is the significant
flexibility built into the formalism.  It can be used to calibrate
even non-redundant arrays as long as they are sufficiently densely
packed.  With further development, it should also be a useful tool in
reconstructing array properties, and carrying out bandpass and
polarization calibrations.  We have provided a single-frequency
reference implementation which we anticipate updating in the future as
more features are supported.

\section{Acknowledgements}

This work was partially supported by National Research Foundation
grants 98957 and 98772.  We thank 
James Aguirre, Josh Dillon, Adrian Liu, Sphesihle Makhathini, Laura
Newburgh, Kendrick Smith, and Keith Vanderlinde for useful
discussions.  Some calculations were carried out on the UKZN hippo
cluster.  

\appendix
\section{Assumption of Gaussianity}

The mathematical foundation of correlation calibration relies on
treating the sky as a Gaussian random field.  The real sky is
obviously not a Gaussian random field, so questions about the
importance and impact of this assumption naturally arise.  This
appendix qualitatively discusses how real images of the sky typically
deviate from Gaussianity, how they impact calibration, and what sorts
of effects would cause difficulties in calibration.

In general, the main way non-Gaussianity appears is in phase
correlations between different modes.  Standard demonstrations{\footnote{{\textit{e.g.}}
http://www.imagemagick.org/Usage/fourier/}} show that the phases
contain most of the information we typically think of as an image.
Redundant calibration explicitly ignores phase correlations between
nominally non-redundant visibilities, and so is intrinsically
insensitive to phase non-Gaussianities since changes to sky phases
leave $\chi^2$ identically unchanged.  So, one can clearly calibrate
in the presence of phase correlations.  If the correlations are even
partly known, then calibration can be improved.  Correlation
calibration takes advantage of exactly this fact when incorporating
sources with known positions.  Generically, the more phase information
known about the sky and included in the likelihood, the better the
calibration, but unmodelled phase correlations will not introduce
biases into the calibration solution.

Real fields can also have non-Gaussian amplitude distributions.  Since
the likelihood depends quadratically on the data, odd moments like the
skewness and noltosis do not affect calibration.  Even moments could
potentially cause problems.  One extreme limit of this is visibilities
dominated by a very small number of point sources, which in general
will be platykurtic (low kurtosis).  The simulations from Section
\ref{sims} would be sensitive to this but show no ill effects.  The
limit of extreme leptokurtosis is more concerning.  Consider the case
of two nearly identical baselines with compact support.  Corrcal
relies on the fact that these should measure very similar visibilties.
However an arbitrarily large plane wave on the sky that one of the
baseline sees but is out of the region of support of the other would
break this assumption.  Fortunately, this is an extremely contrived
case that does not reflect what is seen - foregrounds aren't (nearly)
pure individual plane waves.  Note that even in the case where there
is significant power in a very constricted wavelength range but
isotropic in angle would be correctly handled as long as the power
spectrum used in calculating the covariances reflected this.

Finally, the input sky power spectrum will in general not be perfectly
known and can be spatially varying.  The effects of these errors are also mild, at least in the
case where the errors are smoothly varying with angular scale.  Under
the approximation that the power spectrum is scaled by a constant
value $\alpha$ within a redundant block, one can estimate the effects
using Equation \ref{eq:chisq}.  If one rescales the noise variances by
the same factor, then $\alpha$ factors out of the matrix inverse, and
the overall $\chi^2$ for that redundant block gets scaled by
$\alpha^{-1}$.  In the signal-dominated limit, this corresponds to
misweighting the redundant block(s) with mis-estimated power spectra
by the same $\alpha^{-1}$, and so does not bias the results.  In the
purely noise-dominated limit, the factors of $\alpha$ cancel, and
$\chi^2$ is unchanged.  Intermediate regimes will misweight the
noise/$\chi^2$ by values between 1 and $\alpha$, but again will not
bias the calibration.  As mentioned previously, if unmodelled large
and sharp features existed in the true power spectrum, they could
negatively impact the calibration, but such features have not been
seen and are not expected.

\bibliography{corrcal2}{}
\bibliographystyle{apj}

\end{document}